\documentclass[aps,amsmath,amssymb,superscriptaddress,reprint]{revtex4-1}
\usepackage{dcolumn}
\usepackage{bm}
\usepackage{amsmath}
\usepackage{amssymb}
\usepackage{amsfonts}
\usepackage{graphicx}
\usepackage{fancyhdr}
\usepackage{epstopdf}
\DeclareGraphicsExtensions{.eps}

\usepackage{xcolor}
\usepackage[top=0.75in, bottom=0.75in, left=0.75in, right=0.75in]{geometry}
\fancypagestyle{firststyle}
{
   \fancyhf{}
   \fancyhead[L]{\textbf{\emph{Submitted to the Journal of the Royal Society Interface (2014)}}}
}
\begin{document}
\title{The Organization and Control of an Evolving Interdependent Population}

\author{Dervis C. Vural}
\thanks{Corresponding Author}
\email{dvural@nd.edu}
\affiliation{Department of Physics, University of Notre Dame}
\author{Alexander Isakov}
\thanks{Co-first author}
\affiliation{Department of Physics, Harvard University}
\author{L. Mahadevan}
\thanks{Corresponding Author}
\email{lm@deas.harvard.edu}
\affiliation{Department of Physics, Harvard University}
\affiliation{School of Engineering and Applied Sciences, Department of Organismic and Evolutionary Biology, Harvard University}

\begin{abstract}  Starting with Darwin, biologists have asked how populations evolve from a low fitness state that is evolutionarily stable to a high fitness state that is not. Specifically of interest is the emergence of cooperation and multicellularity where the fitness of individuals often appears in conflict with that of the population. Theories of social evolution and evolutionary game theory have produced a number of fruitful results employing two-state two-body frameworks. In this study we depart from this tradition and instead consider a multi-player, multi-state evolutionary game, in which the fitness of an agent is determined by its relationship to an arbitrary number of other agents. We show that populations organize themselves in one of four distinct phases of interdependence depending on one parameter, selection strength. Some of these phases involve the formation of specialized large-scale structures. We then describe how the evolution of independence can be manipulated through various external perturbations.
 \end{abstract}

\maketitle
\thispagestyle{firststyle}

\maketitle







\section{Introduction}

Cooperative behavior, as exemplified by multicellular life, seems to have evolved at least 25 times independently - once for plants, once or twice for animals, once for brown algae, and possibly several times for fungi, slime molds, and red algae \cite{grosberg}. On shorter time-scales, the social composition of eukaryotes such as \emph{S. cerevisiae},  and biofilm forming bacteria such as \emph{P. aeruginosa} can dramatically change in a brief period \cite{ratcliff, gore, jiricny,diggle}. In a related context, tumor formation is a rare example of the transition, taking place in the reverse direction, from a multicellular to an essentially unicellular lifestyle. Interestingly, cancer cells end up cooperating by collectively secreting angiogenic factors, and it seems possible, at least in principle, that there may even be cheaters (i.e. those who do not secrete the growth factors) among this collection of cooperating cheaters \cite{axelrod,nagy}.

Evolutionary game theory provides excellent insight into how altruistic and cooperative behavior can emerge to maximize the fitness of the group despite the apparent fitness advantage of cheating individuals \cite{maynard,nowak1}. In the context of evolution of cooperation, these models typically investigate the outcome of repeated runs of the prisoner's dilemma between pairs of agents that have two strategies, cheating and cooperating. Variants of the model include structured interactions, coupled populations, coevolution, stored reputation, punishment, and preferential or random partner choosing \cite{west,roca, szabo,perc,rand,perc1, gross,perc-coev,wu}.

However, real life is more complicated in a number of ways. First, many actual games are massively multiplayer \cite{broom,connor,archetti,gokhale}. The fitness of an organism may depend on its simultaneous relationship with multiple players. Second, biology allows for a much larger variety of internal states beyond cooperating or defecting. For example, the genetic makeup of an organism may be suitable for cooperation with only an exclusive few, while some organisms may be incapable of defecting or cooperating all together. Third, real social evolution leads to highly organized dependence structures beyond the homogeneous mixtures or aggregates of cooperator-defector states that we often see in the majority of evolutionary game models. From biochemical to societal scales, life organizes itself in highly complex arrangements of cliques, communities, cycles and hierarchies.

Without compromising the simplicity and tractability offered by traditional evolutionary game theory, here we propose an evolutionary model in which the fitness of an agent is determined, not by the outcome of a two-player two-state game, but instead a multi-player multi-state one. Thus, our focus is not the cooperator / defector ratio in the population, but rather the large-scale structure of all exchanges; i.e. the \emph{interdependence} between agents or groups of agents within a genetically heterogeneous population. We ask how independent agents become interdependent through the simple laws of evolution, whether positive selection is a necessary or a sufficient condition for the formation of interdependence, what kinds of interdependent structures are stable / unstable, and how these structures and processes depend on evolutionary parameters.
\begin{figure}
\includegraphics[scale=0.65]{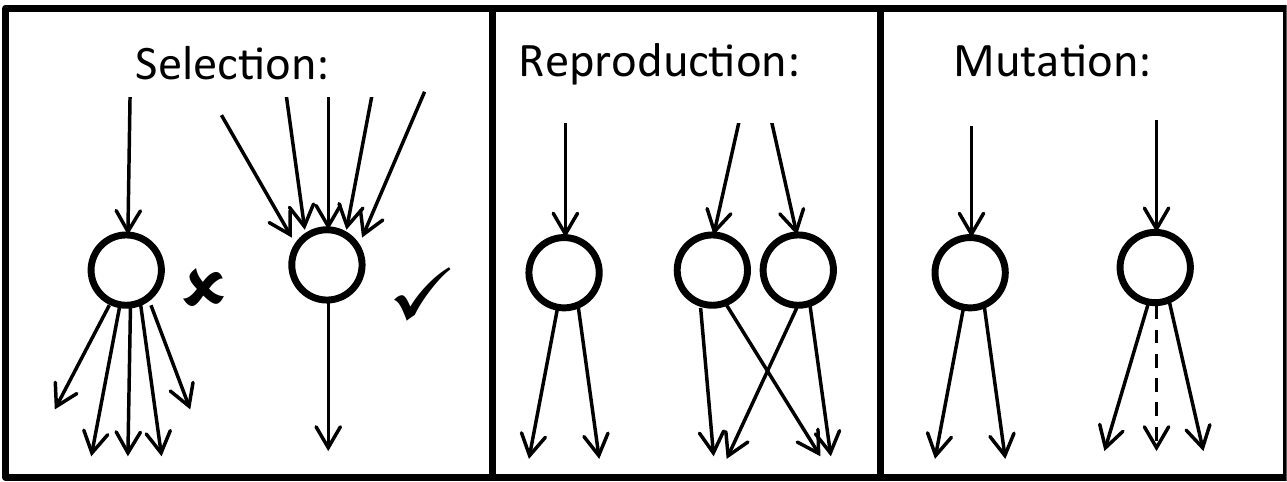}
\caption{{\bf Schematic of the model}. The evolutionary algorithm is carried out by assuming that (1) fitness of a node $\omega(x_i)$ is a monotonically increasing function of edge influx   $x_i=b n_{i, \mathrm{in}}-c n_{i, \mathrm{out}}$, where $n_{i, \mathrm{in}}$ and $n_{i, \mathrm{out}}$ are the number of in and out edges for node $i$.  (2) reproduction preserves all in-out relationships with $r$ fittest nodes replacing the least fittest nodes, and (3) a small number of edges are randomly added/removed every generation, with probability $p \ll 1$.}
\end{figure}
Accordingly, the present model offers a clear framework for classifying and categorizing different regimes of interdependence, as well as allowing for careful control of evolutionary parameters that may be influencing recent non-intuitive empirical outcomes \cite{kohler}. We determine which kinds of external perturbations promote anti-sociality (e.g. in order to eradicate biofilms) and which other kinds can inhibit anti-sociality (e.g. as to suppress or reverse tumor growth) by simulating the introduction of selfish/altruistic strains into a population or the administration of anti-sociality/sociality promoting drugs. We evaluate the success rate of these evolutionary interventions as a function of the original population structure, drug dose, fraction of drug-resistant agents, and reproduction speed of the target species.
\begin{figure*}
\hspace{-0.1in}\includegraphics[scale=0.32]{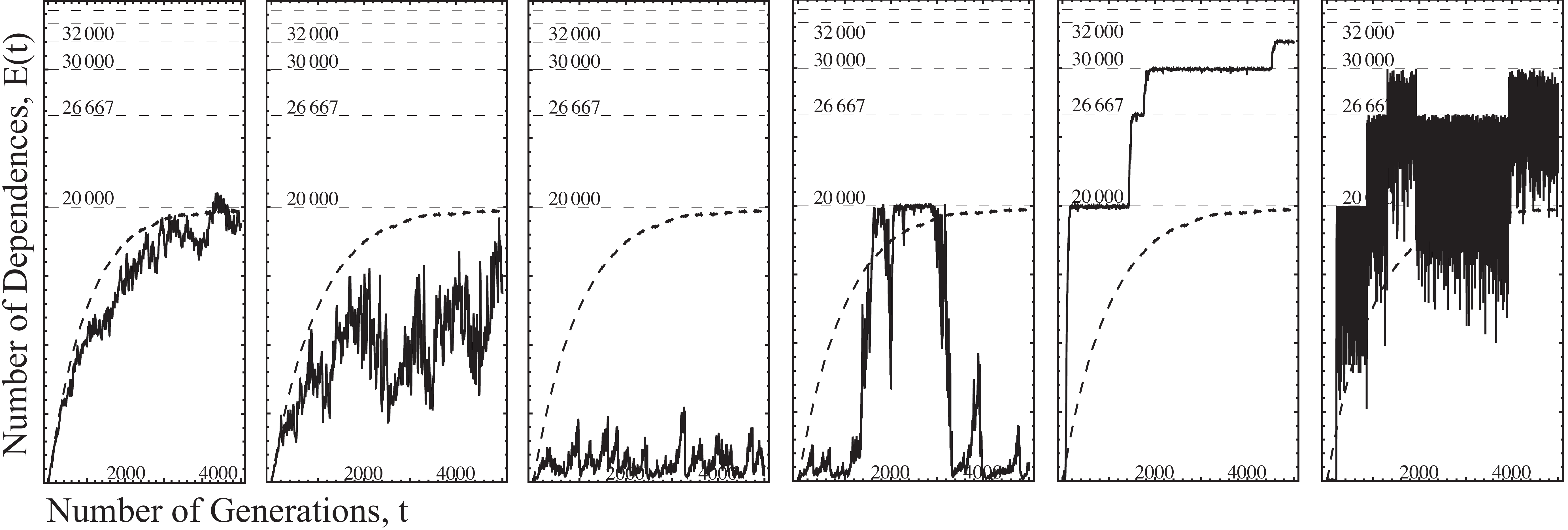}
\caption{{\bf Regimes of constructive evolution ($\beta>1$)}. Plotted is the number of edges $E(t)$ as a function of time (generations) $t$, as the selection strength is increased from left to right $r=2,4,9,10,15,100$ while mutation probability is kept constant $N^2p=20$. The dashed straight lines indicate the stable number of edges corresponding to an integer number $k$ of equal-sized ``bunches'', $E= N^2\left(1-\frac{1}{k}\right)$. The dashed curved line is the outcome of the fully neutral simulation ($r=0$). For all panels $N=200$, $\beta=1.01$.}
\end{figure*}

\section{Model}
Our multi-state multi-player game can be best visualized as a network of $N\gg 1$ agents, connected by directional edges. An edge from $A$ to $B$ indicates that $A$ contributes to the fitness of $B$ at the cost of its own. Unlike the typical evolutionary game theoretic models where the state of a player $i$ is binary (cooperator / cheater), here the player states $\psi_i$ are characterized by high-dimensional vectors, i.e. $\psi_i=\{x_1,x_2,\ldots,x_N\}$ with $x_j\in\{0,1\}$ indicating whether $i$ provides a fitness benefit to $j$. The evolutionary dynamics is governed by the following assumptions (Fig[1]):

{\bf(1)} The fitness $\omega(x_i)$ of a node $i$ is assumed to be a monotonically increasing function of received net benefit $x_i=b n_{i, \mathrm{in}}-c n_{i, \mathrm{out}}$, where $n_{i, \mathrm{in}}$ and $n_{i, \mathrm{out}}$ are the number of in and out edges for node $i$. The parameter $\beta=b/c$ quantifies the benefit of an edge (to the receiver) relative to its cost (to the provider).

{\bf(2)} Every generation, the $r$ most fit nodes produce offspring that replace the $r$ least fit nodes. Reproduction preserves all edge relationships of the parent, i.e. parents and offspring connect to the same agents.

{\bf(3)} There is a small mutation probability $p$ per generation with which edges are added/removed randomly.

In other words, we have a fitness-based selection rule keeping the number of agents $N$ constant. Our simulation code is available as an electronic supplement.

Our model has four parameters, kept constant throughout the course of evolution: Population size $N$, mutation probability $p$, number selected for replacement $r$, and the relative benefit $\beta$. For every run we keep track of the total number of edges $E(t)$ as a function of generation number $t$. $E(t)$ is a measure of the interdependence of the population as well as the average fitness (the latter follows from $\langle\omega\rangle=\sum_i\omega_i/N=(\beta-1)E(t)/N$, which can be positive or negative depending on the value of $\beta$). In addition, we study the community structures and genetic composition within the population, which are defined in terms of the connectivity matrix $C$ of the network. We use the convention that $C_{ij}=1$ if $j$ depends on $i$, and $0$ otherwise, and represent these by black and white pixels in array plots. Our simulations were run for $N=200$, partly due to computational constraints. Although this might appear small, we note that the relevant degree of freedom here is the number of edge slots $N^2 = 4\times10^4$, ensuring that the evolutionary transitions we report are not accidental fluctuations.

\section{Results}
The evolutionary dynamics and final interdependence states depend on the values of $\beta$ and relative selection pressure $r/m$, where $m=N^2p$ is the expected number of mutations per generation (which is equal to the number of mutants if $p\ll1$). We can divide the parameter space into four regimes: Neutral constructive ($r\ll m, \beta>1$), selective constructive ($\mathcal{O}[r]\sim\mathcal{O}[m], \beta>1$)), neutral destructive ($r\ll m, \beta<1$) and selective destructive ($\mathcal{O}[r]\sim\mathcal{O}[m],\beta<1$). In all cases, the formation of an edge is beneficial for one of the nodes and deleterious for the other. However the value of $\beta$ determines the change in average fitness per edge, $d\langle\omega\rangle/dE=c(\beta-1)/N$, and therefore one intuitively expects $E(t)$ to decrease for $\beta<1$ and increase for $\beta>1$ as long as selection strength is finite. We will see that this expectation will not always be satisfied, particularly when $\beta>1$.

Let us start with the straightforward case of $\beta<1$. Here the formation of an edge is more deleterious to its originator than beneficial to its target, and the fitness of the population changes by $b-c<0$ per edge. Thus, in the long run, only if the selection is weak ($r\ll m$) can such deleterious edges accumulate, and we get a random interdependence network with $E=N^2/2$. As expected, increasing $r/m$ causes the network to become sparse and fragmented, and all structure vanishes as $\mathcal{O}[r]\sim\mathcal{O}[m]$.

We now overview the constructive regime $\beta>1$, which produces distinct phases of complexity  (Fig[2], panels 1-6). The long term behavior of $E(t)$, which can be viewed as a proxy for average fitness as well as interdependence and complexity, depends non-monotonically on selective pressure $r/m$: For small values of $r/m$, the asymptotic value $E(t\to\infty)$ decreases with $r/m$ (Fig[2], panels 1-3). However if $r/m$ exceeds a critical point we see sudden transitions between well-defined discrete levels (Fig[2], panels 4-6). As selective pressure is increased further, we see increasingly larger fluctuations around these levels.

We describe these asymptotic states in more detail by connectivity matrices (Fig[3], top row) and phylogenetic trees (Fig[3] bottom row) for varying levels of $r/m$. The phylogenetic trees are obtained by quantifying the similarity distance $D_{ij}=\sum_{k}|C_{ik}-C_{jk}|+\sum_{k}|C_{ki}-C_{kj}|$ between all pairs of nodes $i$ and $j$. In other words, if $i$ and $j$ receive from and provide to the same nodes, they are considered to be genetically related, consistent with our reproduction rule (cf. Fig[1]). 

While the destructive $\beta<1$ regime produces either random or sparse networks, the constructive regime $\beta>1$ can be summarized in terms of a sequence of complex phases governed by the value of $r/m$ (Fig[4]): A transition from cooperation to competition between individuals (Fig[2] panels 1-3) is followed by unstable interactions between individuals and ``bunches'' (Fig[2] panel 4), followed by a transition from cooperation to competition between ``bunches'' (Fig[2] panels 5-6). We define a \emph{bunch} to be the opposite of a graph theoretical community; a group of nodes that form denser connections towards other groups, than they do within (cf. latter two panels in Fig[3]).  Dense outwards connections and sparse intra-connections are the key qualitative characteristic of a highly specialized system. For example, nearly all energy spent by a heart muscle cell is directed at serving other tissues. The same holds true in a specialized society, e.g. a lawyer dedicates most of her effort defending non-lawyers. The interdependence structures we report in the latter two panels of Fig[3], conform to these biological and social examples of specialization.

We now move towards an understanding of the \emph{control and manipulation} of the evolution of interdependence, which is now experimentally possible (albeit with mixed success) in biomedical and ecological settings. For example, the sociality of \emph{P. aeruginosa} can be manipulated by drugs that suppress the microbe's production of a common good (iron scavenging siderophores). Since the microbes that are resistant to the drug will altruistically continue to produce the expensive siderophores, they are taken over by their selfish counterparts affected by the drug \cite{mellbye,sandoz,diggle}. As a result, the iron-deficient population can be easily annihilated by the host's immune system \cite{xavier}. Note that the evolutionary fate of the drug-resistant group would have been the opposite, had the drug been an antibiotic instead of a quorum blocker. On the other hand, there have also been experiments yielding the \emph{exact opposite} outcome, where the drug aggravates the infection instead of impairing it, presumably by leveling the relative advantage of cheaters \cite{kohler}. We will use our model to quantify these mixed outcomes. Social evolution is complex, and its manipulation and control requires a detailed quantitative understanding of the evolutionary outcomes of varying initial states and system parameters.

To manipulate the sociality of a highly interdependent, $\beta>1$ population, we start with an initial network that has a given community structure, select a fraction $\eta$ of the population, and block a fraction $\gamma$ of their outgoing connections of those that are selected. Following this perturbation, we track the evolution of the network and check if our perturbation causes the entire population to lose all connections (which, for $\beta>1$, amounts to minimal fitness). If $E(t)$ drops to and remains at zero we count it as a success and we determine the fraction of successes for every parameter value.  If $\eta \ll 1$, the perturbation can be interpreted as an external introduction of a new strain/species, or a novel mutation which introduces a very small number of selfish individuals in the population. If $\eta \simeq 1$ the perturbation can be interpreted as a drug or event that inflicts nearly everyone, such as the quorum blocker discussed earlier. Accordingly, the quantity $\gamma$ can be interpreted as the \emph{dose} of the drug, or the degree of ``selfishness'' of the newly introduced species/strain. 

Fig[5] displays the dependence of success as a function of $\eta$ (empty vs. closed plot markers correspond to $\eta=2\%$ and $98\%$), initial population structure $k$ (quantifying the number of bunches) and $\gamma$. We consider fast reproducing and slow reproducing species separately, shown in the left and right panels respectively. 

\begin{figure*}
\hspace{-0.05in}\includegraphics[scale=0.2]{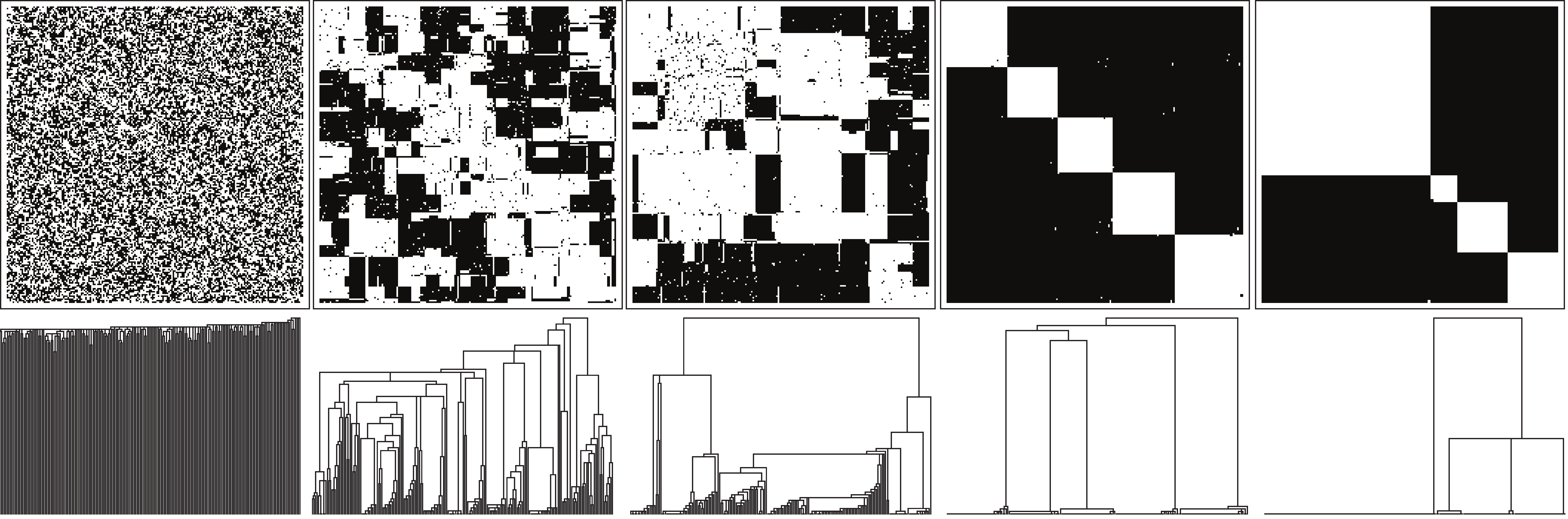}
\caption{{\bf Interdependence and genetic composition}. Connectivity matrices (top) and their respective phylogenetic trees (bottom) show the dramatic difference in the final organization of the population caused by varying selective strength (left to right, $r=0,2,4,15,100$) in the constructive regime ($\beta>1$). The connectivity matrix element $C_{ij}$ is marked by black if individual $i$ provides fitness to $j$, and left white if there is no exchange. The tree linkages in the bottom are formed according to smallest intercluster dissimilarity, defined by the $L_1$ norm. For all panels $N=200$, $N^{2}p=20$, $\beta=1.01$ is kept constant. We see the onset of ``bunch'' (anti-community) formation even in the weak selection limit (compare panel 1 to 2). The number and definition of bunches increases with higher selection strength (panel 4). In the strong selection limit bunches compete with each other, leading to size heterogeneity (panel 5).}
\end{figure*}
It is important to distinguish between two very different mechanisms that bring the population back to its pre-perturbed state. The first is determined by the time required for interdependence to evolve anew from $E=0$. The original factors causing the establishment of cooperation in the first place is present regardless of our perturbation, and the effect of even the strongest drug ($\eta=1, \gamma=1$) is to simply reset the evolutionary clock. The second mechanism is evolution through the repopulation of the drug-resistant fraction, which happens much faster, on reproductive time-scales. To clearly distinguish between these two mechanisms, we set $p=0$ in Fig[5]; using a finite $p$ scales down all the success rates, but does not otherwise change the qualitative dependence on $\eta$, $\gamma$ and $k$.

We observe a number of interesting features in the response of the population to external perturbations. For $\eta\simeq1$ we find a non-monotonic dependence of success rate to $\beta$ for populations with few bunches: For slow reproducing populations a moderate dose works as well as, or better than, a strong one. For larger number of bunches, and faster reproduction rates the non-monotonicity vanishes: The stronger the dose, the better the outcome. A second remarkable outcome is the degree to which a few individuals can make a difference: Targeting $\eta=2\%$ of the population is as effective as targeting $\eta=98\%$ of the population provided the drug has a high enough dose. This is because few selfish individuals, as is the case in tumors or invasive species, can devastate an entire population. Finally, we observe a very strong dependence of the success rate on the initial community structure. With increasing $k$ and $r$ this difference vanishes. 
\begin{figure}
\hspace{-0.3in}\includegraphics[scale=0.43]{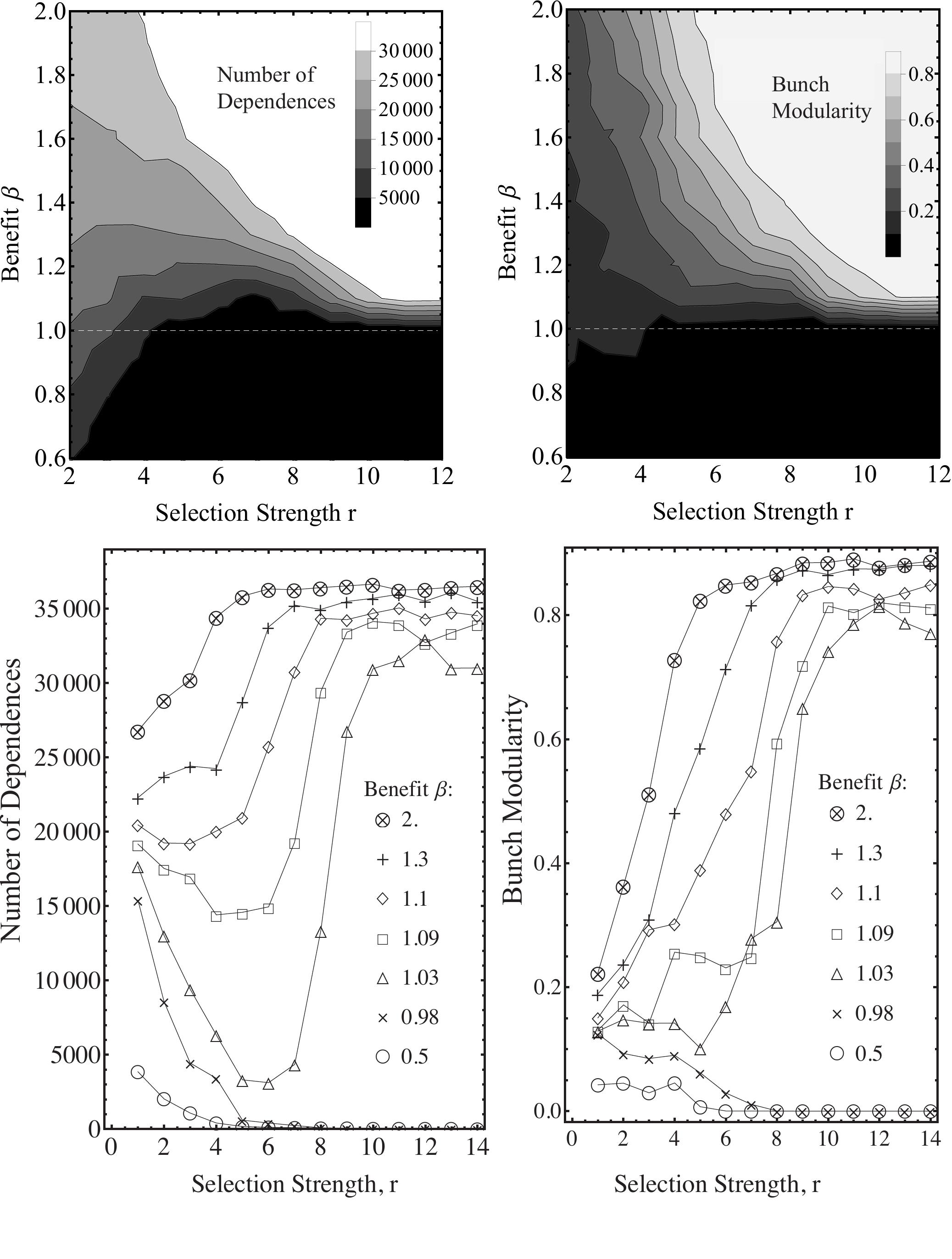}
\caption{{\bf Phases of interdependence}. The phase diagrams (top row) displays the asymptotic dependence number $E$ (left) or the bunch modularity (right) as a function of relevant system parameters. Traversing both phase diagrams (bottom row) in the horizontal direction clarifies the phase profile. Since the some of the phases are highly dynamic, the $E$ (left-top) is taken to be the minimum number of dependences in a large time-window in the long-time limit.} 
\end{figure}
\section{Discussion}
The phase diagram for the evolution of interdependence is shown in Fig[4] and compactly summarizes our results. Panel 1 shows the number of dependences, while Panel 2 quantifies their structure through ``bunch modularity''. We define the latter by exchanging $1\leftrightarrow0$ in the connectivity matrix and determining the community modularity. Throughout, we will discuss and label separate phases of interdependence by letters A-D.

\subsection{Cooperation Between Individuals} In the neutral regime ($r\ll m$) additions and deletions of edges are equally likely. Thus, if the population starts fully independent, $E(t)$ increases until the network is fully randomized with $N^2/2$ edges. This increase is statistically irreversible, and is the analogue of the scenario described in \cite{stoltzfus,gray,vural}. In the neutral regime individuals have high fitness due to the benefits of indirect reciprocity; interdependence emerges not due to higher fitness but simply due to higher likelihood. Fig[1a] shows the dynamics and final outcome of nearly-neutral evolution. Despite following a similar trajectory to fully-neutral evolution ($E(t)\sim N^2(1-e^{-2pNt})$, indicated by the dashed curve in all panels of Fig[2]), the connectivity matrix shows the onset of community formation; the interdependence structure and genetic composition of the population is far from random (compare Fig[3a, 3b]).

As the selection strength is increased ($r<m$ but not $r\ll m$) the fluctuations in $E(t)$ are amplified. This is caused by the random formation of nodes for which the number of in-edges are different than out-edges. However the system is self stabilizing (Fig[2], panels 1,2); e.g. when the fit defectors reproduce, they typically replace their unfit providers, which in turn reduces their own fitness. Consequently they are taken over by the fair and fit nodes that dominate the $r\ll m$ population. In Fig[3] panel 3, two large reciprocating groups can be distinguished. They are taken advantage by smaller scale opportunistic sub-populations. It is also possible to see a smaller sub-cooperative group sustaining itself within a larger cooperative group.

\subsection{Competition Between Individuals} As $r$ approaches $r_{c}\sim m/2$ from below, the fluctuations in $E(t)$ become comparable to $E(t)$ itself. Here the selective competition is just high enough to allow for small cooperative communities to form and grow at a rate much higher than random chance, but also high enough for cheaters to spread over their providers in one step, beyond recovery (Fig[1], panel 3). Although regime A and B have similar destabilizing factors, their re-stabilization is very different. The drops in $E(t)$ in regime A can recover through \emph{re-population}, over time scales $\sim1/r$. In contrast regime B exhibits system-size losses from which the only way to recover is \emph{re-mutation}, over longer time scales determined by $\sim1/m^2$, as the smallest cooperative group requires two mutations. 

Comparing A to B reveals that higher selection strength in this case leads to lower fitness; had one mixed the stronger-selected population B with the weaker-selected one A, the former would be driven to extinction. The behavior of B is similar to that expected in a classical prisoner's dilemma, which emerges from our model as a special case - survival of the fittest produces the globally least fit outcome. 

\subsection{Formation of Specialized Bunches}
As $r$ is increased above the critical point $r_c\sim m/2$ higher structures start to form. While the connectivity matrix $C$ is sparse and random for $r$ just below $r_c$, we start seeing metastable bunches at $r>r_c$, the number and stability of which increases with $r$. The sudden jump in the edge number we see in Fig[3], panels 4-5 is analogous to that found in \cite{jain}.

For a very large window of selective strength $\mathcal{O}[m]\sim r<\mathcal{O}[N]$, we see that the system can only maintain certain discrete values of $E$. These are the stable configurations corresponding to an integer number of equal sized bunches ($k$) given by the relation $E_k=N^2\left(1-\frac{1}{k}\right)$, $k=0,1,2,\ldots,k_{\mathrm{max}}$ (Fig[2] panel 4). The maximum number of bunches $k_{\mathrm{max}}$ is determined by the mutation rate, $k_{\mathrm{max}}\sim N/m$ (i.e. so that in steady state there is one mutation per bunch per time step). However, we have observed $k$ transiently increasing to $50\%$ higher than this value. Note that the degree of interdependence (and hence the average fitness) in the strong selection limit well exceeds that in the weak selection limit. 

It is interesting that in the limit $r,m\sim1$, we see structures more complex than bunches. These include hierarchies (smaller bunches within a bunch), cycles (3 or more groups providing to one other), and hierarchies of cycles (cycles within a cycle). In this limit the dynamics of $E(t)$ still exhibits discrete steps similar to Fig[2] panel 5, however with more possible metastable plateaus corresponding to unequal sized matrix blocks.

\subsection{Competition Between Bunches} With increasing $r$ the fluctuation in the number of edges around the stable $k$ starts increasing, and we see destructive competition similar to that near the phase boundary of B; however, now the competition is between the bunches rather than the individuals, which creates significant size differences between them. These fluctuations can lead to one bunch replacing another, causing $E_k$ to make large transitions between different values of $k$. Despite the apparent noise (Fig[2] panel 6) the dependence structure remains in a highly ordered state with high reciprocity. As $r$ is increased further we see that competition between bunches cause fluctuations comparable to the size of bunches. i.e. a small bunch can increase in size by spreading over others until another metastable structure re-evolves.

\begin{figure}
\hspace{-0.13in}\includegraphics[scale=0.31]{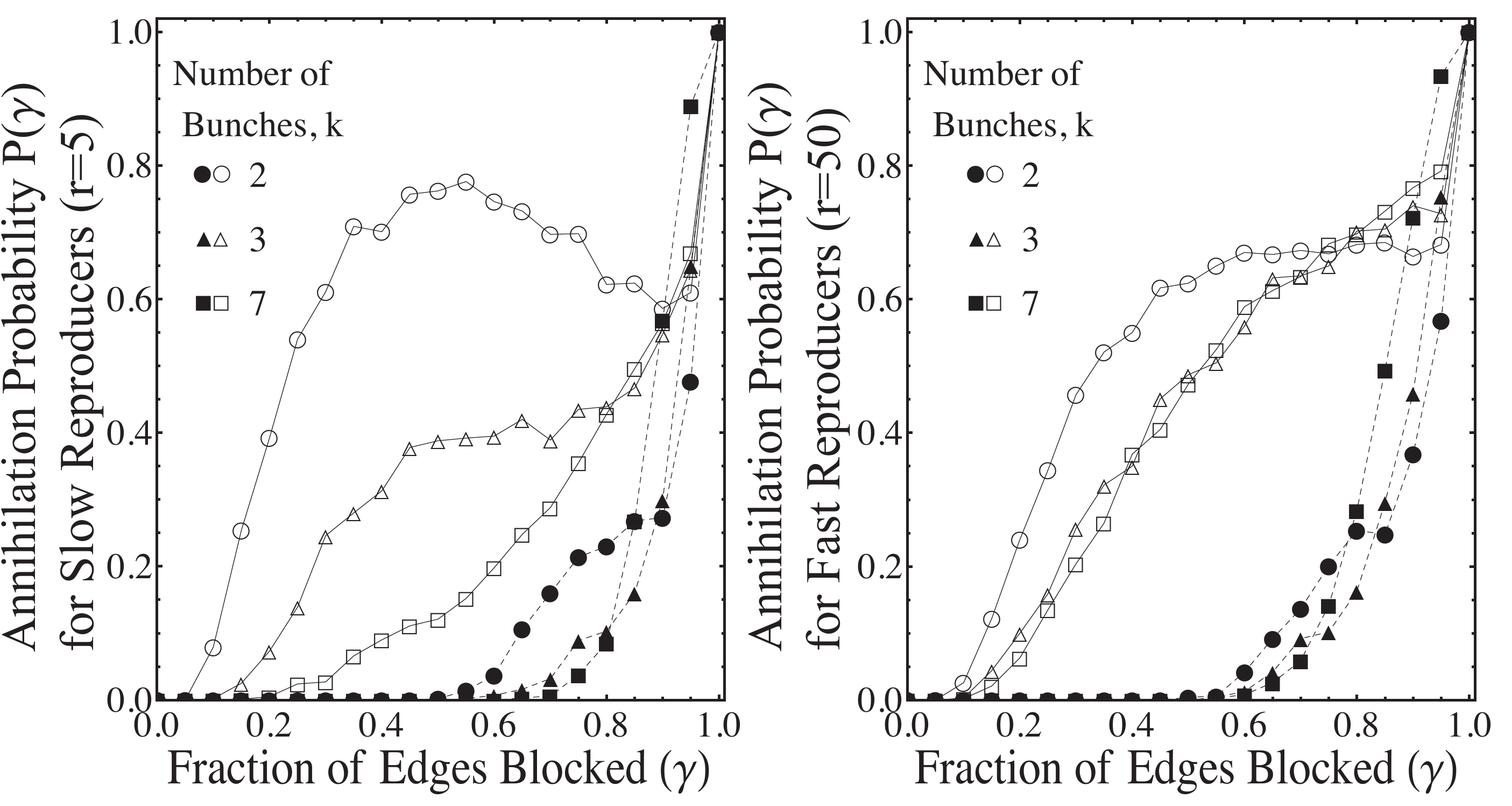}
\caption{{\bf Manipulation and Control of Social evolution}. Of the entire population (N=200) we mark $\eta=2\%$ (empty markers) and $\eta=98\%$ (filled markers) of the nodes as susceptible to our perturbation. We then block $\gamma$ (horizontal axis) of the out edges of the susceptible population, and determine the probability that the strong cooperating population collapses into a weak non-cooperating one (i.e to $E=0$). To clearly distinguish between the spread of the resistant subpopulation $\eta$, and novel mutations that occur after our perturbation, we set $p=0$ (finite $p$ merely adds noise to the curves). We observe that the effect of the drug strongly depends on the community structure for slow reproducing species (left, $r=15$) but not for fast reproducing ones (right, $r=60$). The dashed curve shows how devastating even a small number of antisocial individuals be for the whole population, and suggests treatments where individual cells are targeted. The strong dependence of annihilation probability on $k$ for small $k$ (left panel) might explain why quorum blockers are effective against some biofilms forming bacteria but not others. Interestingly, for $k=2$, $P(\gamma=0.5)$ is significantly higher than $P(\gamma=0.95)$, suggesting that for small number of cooperative bunches smaller perturbations may be more effective than larger ones.}
\end{figure} 
\section{Conclusions} 

We have constructed a simple model that allows us to study the co-evolution of self-replicating interdependent structures, and reported multiple evolutionary transitions as $\beta$ and $r$ are varied. This model is quite general and has very few assumptions - the fitness function is only assumed to be an arbitrary increasing function of $x$ and there are only two relevant parameters governing the dynamics (selection strength $r/m$ and relative benefit $b/c$) since the population size $N$ does not make a qualitative difference as long as both $Np$ and $r$ are much smaller than $N$. Furthermore, the value $\beta$ does not make a qualitative difference apart from whether it is larger or smaller than unity, and no \emph{quantitative} difference if $|1-\beta|$ is smaller than $1/N$. Unlike the typical simplified models of evolutionary game theory we do not assume that an individual's behavior is the same towards all others (although some individuals can end up in a state where they give to all and receive from all). In this respect the states allowed in this work is a generalization of the two-state models common in the literature. Thus, we hope that our model can serve as a guiding framework for understanding the emergence of sociality.

Even in this simple case, we observe a number of surprising and important phenomena. First, we report that even the weakest selection strengths ($m\gg r$) can produce interdependence structures that are far from random. Thus, assumptions regarding ``random interdependence'' invoked by neutral evolutionary arguments may be too strong \cite{stoltzfus,gray,vural}. Our second observation is the natural emergence of specialized bunches and multi scale structures from the simple laws of evolutionary dynamics. As we probe the response of the system to various selection strengths, we see regimes of random interdependence, competition between nodes, cooperation between nodes (bunches) and competition between bunches.

Thirdly we report that the regime $\beta>1,~r>0$ does not ensure complex interdependence. There exists a ``dead zone'' within the constructive regime (Fig[2c]) due to the competition between agents. This non-monotonic dependence of $E$ on selection strength can have important implications in medicine. For example, biofilm populations may be induced into a less virulent non-cooperative state by decreasing the selective pressure, so that a cooperative film behaving as Fig[2e] evolves into an intermediate non-cooperative state behaving as Fig[2c]. This may be experimentally verified in \emph{P. aeruginosa} by increasing the available iron while keeping their population constant by limiting their carbon source.

Another remarkable result is the non-monotonic connection between anti-social drug dose and the successful annihilation of cooperativeness. Indeed, the model exhibits a a ``contagion'' effect which allows the manipulation of a few individuals to have population-wide effects. It has been noted that introducing several selfish mutants (or using an anti-social drug effective on a few individuals) may be far more effective than manipulating an entire population \cite{kohler} and is consistent with experimental observations \cite{diggle,mellbye,sandoz}.

Finally, when mutation rate is set to zero we observe that the behavior of $E(t)$ resembles that of classical population dynamics. The dynamics between providers and receivers becomes qualitatively similar to that between the predators and prey of a Lotka-Volterra type system. Starting from a randomized connectivity matrix and setting $p=0$, it is common for $E(t)$ to reach a fixed value and oscillate around it. 

Due to its generality and applicability, our model has room for many natural extensions. For example, the distribution of parameters and fitness functions in a more realistic model could include spatial, temporal and individual heterogeneity. Further, the quantities $p$, $b$ and $c$ can be dynamic as they are themselves, to an extent, subject to evolutionary forces. This can lead to a very interesting set of potential future studies exploring connections between interdependence and evolvability/efficiency. Another factor not taken into account here is the possibility of the change in population size due to statistical fluctuations (e.g. due to a time dependent energy input, or infection/predation). Such extensions would be appropriate to address systems in ecology, structured biological population, and provide insight into complicated social trends.
{\bf Acknowledgement:} This work was funded by the Wyss Institute for Biologically Inspired Engineering, the Harvard Kavli Institute for Bio-nano Science and Technology, the MacArthur Foundation, and government support (A.I.) under FA9550-11-C-0028 awarded by the Department of Defense, Air Force Office of Scientific Research, National Defense Science and Engineering Graduate Fellowship, 32CFR168a.

\bibliographystyle{apalike}

\end{document}